\documentclass[twocolumn,nobibnotes,nofootinbib,superscriptaddress]{revtex4}
\usepackage{amsmath,amssymb,bm}
\usepackage{epsfig}
\usepackage{graphicx}
\usepackage{pstricks,pst-node,pst-text,pst-3d}

\newcommand{\beeq}{\begin{eqnarray}}
\newcommand{\eeeq}{\end{eqnarray}}
\newcommand{\be}{\begin{equation}}
\newcommand{\ee}{\end{equation}}
\newcommand{\bea}{\begin{array}}
\newcommand{\eea}{\end{array}}

\newcommand{\eq}{&=&}
\newcommand{\eto}{{\mbox{\textrm e}}}

\def\xp{x_{{I\!\!P}}}

\def\qbar{\overline{q}}
\def\cbar{\overline{c}}
\def\bbar{\overline{b}}
\def\sigmahat{\hat{\sigma}}

\def\half{\textstyle{\frac{1}{2}}}
\def\gev{{\textrm GeV}}
\def\rbo{{\bf r}}
\def\bbo{{\bf b}}

\begin{document}
\title{\bf Dipole model analysis of the  newest diffractive deep inelastic scattering data}

\author{K. Golec-Biernat}\email{golec@ifj.edu.pl}
\affiliation{Institute of Physics, University of Rzesz\'ow, Rzesz\'ow, Poland}
\affiliation{Institute of Nuclear Physics Polish Academy of Sciences, Cracow, Poland}
\author{A. \L{}uszczak}\email{Agnieszka.Luszczak@ifj.edu.pl}
\affiliation{Institute of Nuclear Physics Polish Academy of Sciences, Cracow, Poland}

\begin{abstract}
We analyse the newest diffractive deep inelastic scattering data from the DESY collider HERA with the help of dipole models. We find good agreement with the data on the diffractive structure functions provided the diffractive open charm contribution is taken into account. However, the region of large  diffractive mass (small values of a parameter $\beta$) needs some refinement with the help of an additional gluon radiation.
\end{abstract}
\pacs{}
\keywords{deep inelastic scattering, diffraction, quantum chromodynamics}

\maketitle

\section{Introduction}

Diffractive deep inelastic scattering (DDIS), observed at the DESY collider  HERA (see \cite{Aktas:2006hy,Chekanov:2008cw} and references therein) is one of the most intriguing phenomenons in the electron-proton $(ep)$ collisions. Despite high virtuality of the photonic probe, the
incoming proton scatters intact being separated by a rapidity gap from  a diffractive system, which is additionally formed in the final state. 
The understanding  of these processes based
on quantum chromodynamics (QCD) is the biggest challenge in the area of deep inelastic scattering. In this class of processes, large photon virtuality $Q^2$, which serves as a hard scale, suggests tha one use perturbative QCD with quarks and gluons as basic quanta. On the other hand, softness of the proton side and formation of the rapidity gap touch fundamental problems concerning transition into the nonperturbative domain of QCD. Thus, such important issues like parton saturation, unitarity  and even confinement, are likely to be addressed in the theoretical description of diffractive processes.

The most promising QCD based approach to deep inelastic scattering (DIS) diffraction is formulated in terms of dipole models. 
In these models,  the diffractive, color singlet state is systematically built  from  parton components of the light cone wave function of the virtual photon (see \cite{Wusthoff:1999cr,Hebecker:1999ej}
and references therein). The lowest order  states is formed by  a quark-antiquark pair $(q\qbar)$ while in higher orders
more gluons $g$ and $q\qbar$ pairs are present. In our analysis we will concentrate on two first components, $q\qbar$ and 
$q\qbar g$, since in the configuration space they can be treated as simple, quark or gluon,  color  dipoles.
Their interaction with the proton is described by the dipole scattering   
amplitude $N(x,r,b)$. Here $r$ and $b$ are  two-dimensional vectors 
of transverse separation and impact parameter, respectively, and $x$ is
the Bjorken variable which brings the energy dependence into the dipole models. The main advantage of this approach is the observation that the dipole scattering amplitude can be extracted from the DIS data on fully inclusive quantities, like the
structure functions $F_2$ and $F_L$, based on  some physically motivated form with a few parameters to fit  \cite{Golec-Biernat:1998js,Forshaw:1999uf,Kowalski:2003hm,Iancu:2003ge}. 
Then, it can be  used in  the description of diffractive processes 
\cite{Golec-Biernat:1999qd,Forshaw:1999ny,Forshaw:2004xd,Forshaw:2006np,Marquet:2007nf,Kowalski:2006hc}.
The  form of $N$ which we use in our analysis is motivated by key features of parton saturation in dense partonic systems. The most important one is a saturation scale $Q_s(x)$ \cite{Golec-Biernat:1998js} which can be extracted from the DIS data on the structure function $F_2$. The QCD based motivation for the existence of such a scale is provided by the analysis of the high energy nonlinear evolution equations of   Balitsky and Kovchegov \cite{Balitsky:1995ub, Kovchegov:1999yj,Kovchegov:1999ua,Kovchegov:1999ji}.

In this analysis, we consider two  important parameterisations of the dipole scattering amplitude, 
called Golec-Biernat-Wuesthoff (GBW)  \cite{Golec-Biernat:1998js} and color glass condensate (CGC)  \cite{Soyez:2007kg}, in which parton saturation results are built in.
We present a  precise comparison of the results of the dipole models which
use these parameterisations with  the newest data from HERA on  the diffractive structure functions, obtained by the H1  \cite{Aktas:2006hy} and ZEUS \cite{Chekanov:2008cw,Chekanov:2008fh} Collaborations.
We also make  a comparison with  new data on the diffractive open charm production \cite{Aktas:2006up}.
An analysis of exclusive diffractive processes within the dipole approach was performed in \cite{Kowalski:2006hc}.  
Previous analyses which use parton saturation results, 
like those in \cite{Golec-Biernat:1999qd,Bartels:2002cj,Goncalves:2004dv,Kormilitzin:2007je,Marquet:2007nf}, are based on  less precises diffractive data, and in consequence, they could not address important questions related the precise comparison  presented in this paper.

The comparison we performed prompts us to discuss some subtle points of the dipole
models, mostly related to the $q\qbar g$ component, and connect them to the approach based
on the diffractive parton distributions evolved with the Dokshitzer-Gribov-Lipatov-Altarelli-Parisi (DGLAP) equations. Within the latter approach,  the  diffractive open charm production is particularly interesting
since it is sensitive to a diffractive gluon distribution. However, the accuracy of the existing
data on such a production does not allow one to discriminate between different gluon distributions considered in
our analysis.

The outline of this presentation is the following. In Sec.~\ref{sec:dsfindm} we present basic formulae
of the color dipole approach to diffraction while in Sec.~\ref{sec:dipscatamp} we discuss the two
parameterisations of  the dipole scattering amplitude used in our analysis. In Sec.~\ref{sec:charm}
we perform a comparison of the dipole model results on the diffractive charm production with the HERA
data. A similar comparison for the total diffractive structure functions is presented in Sec.~\ref{sec:f2d}.
In Appendix we derive a formula for the diffractive gluon distribution from dipole models, which is
important for the discussion of the diffractive charm production.

\section{Diffractive structure functions in dipole models}
\label{sec:dsfindm}

\begin{figure}[b]
\vskip -1cm
\centerline{
         \includegraphics[width=5cm]{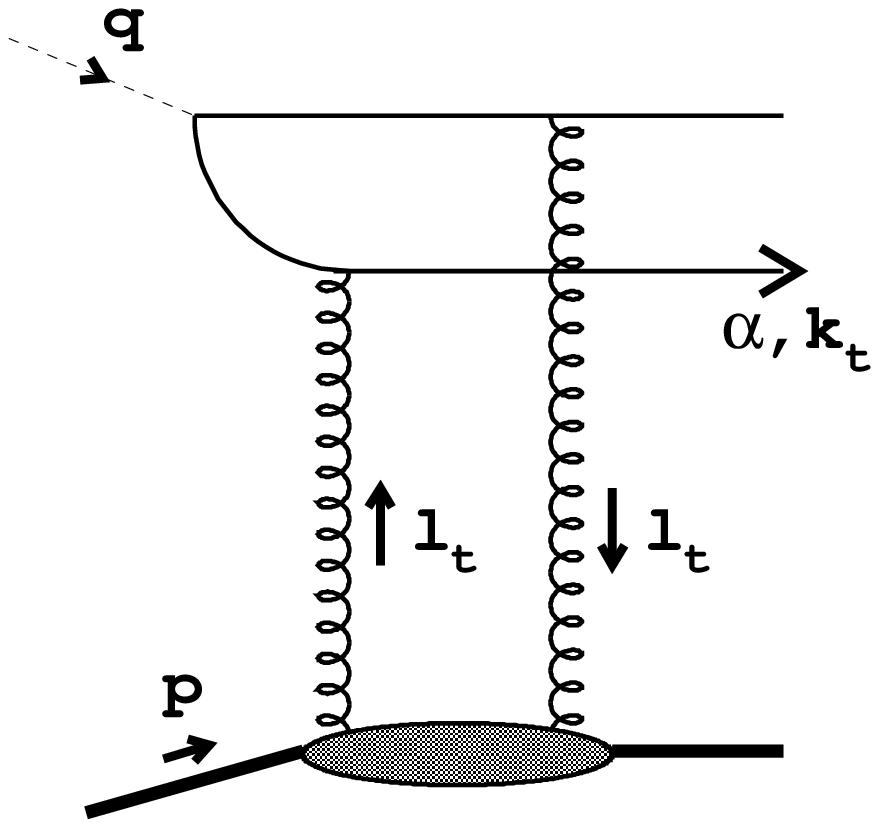}
         \includegraphics[width=5cm]{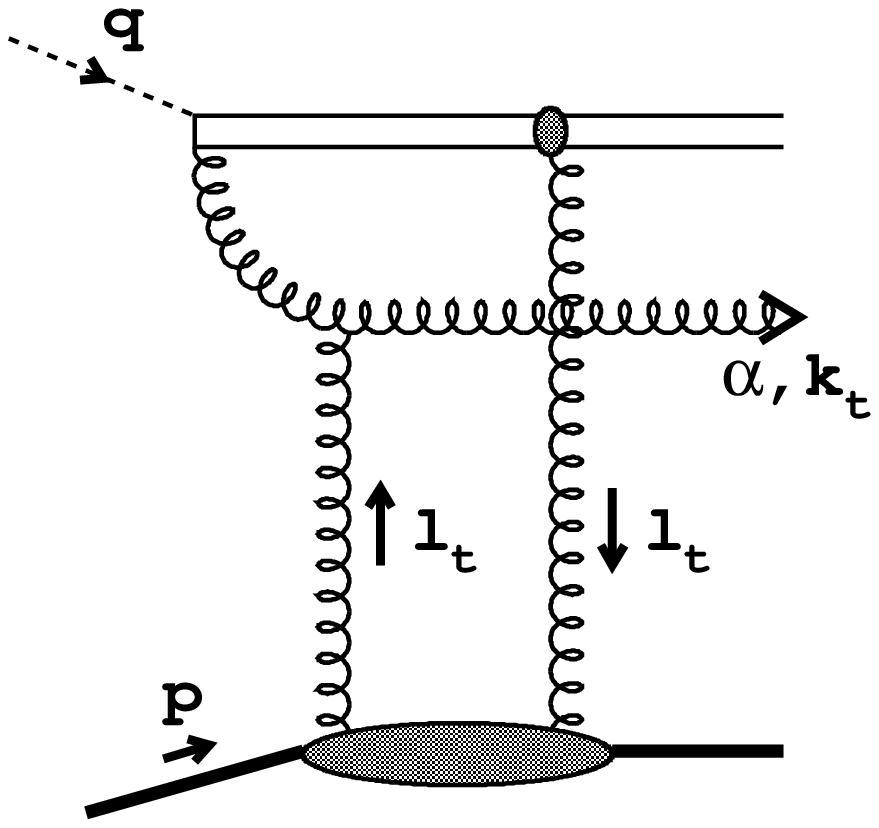}
           }
\caption{The $q\qbar$ i $q\qbar g$ components of the diffractive structure function $F_2^D$.}
\label{fig:components}
\end{figure}

In the dipole approach to DDIS, the diffractive structure function
$F_2^{D}$ is a sum of components corresponding to different diffractive final states produced by a
transversely $(T)$ and longitudinally $(L)$ polarised virtual photon  \cite{Bartels:1998ea}. 
We consider a two component diffractive final state which is built from a $q\qbar$ pair from a transverse and longitudinal photon anda $q\qbar g$ system from a transverse photon, see Fig.~\ref{fig:components}.
Thus,  the structure function is given as a sum
\be\label{eq:1}
F_2^{D}(\xp,\beta,Q^2)=F_T^{(q\qbar)}+F_L^{(q\qbar)}+F_T^{(q\qbar g)}\,,
\ee
where the kinematic variables depend on diffractive mass $M$ and center-of-mass energy of the $\gamma^* p$ system
$W$ through
\be\label{eq:2}
\xp = \frac{M^2+Q^2}{W^2+Q^2}\,,~~~~~~~~~~\beta=\frac{Q^2}{Q^2+M^2}
\ee
while the standard  Bjorken variable $x=\xp\beta$. The dependence of $F_2^D$ on the momentum transfer $t=(p-p^\prime)^2$ is integrated out.
The $q\qbar$ components from  transversely and longitudinally polarised  photons are given by 
\beeq\nonumber
\label{eq:5a}
\xp F_T^{(q\qbar)}\!\!\eq\!\!
\frac{3 Q^4}{64\/\pi^4\beta B_d}\,\sum_f e_f^2\int_{z_{f}}^{1/2}
dz\, z(1-z)
\\
&\times&\left\{
[z^2+(1-z)^2]\,Q^2_f\,\phi_1^2 \,+\,m_f^2\, \phi_0^2
\right\}~~~~~~~~
\\\nonumber
\\\label{eq:5}
\xp F_L^{(q\qbar)}\!\!\eq\!\!
\frac{3 Q^6}{16\/\pi^4\beta B_d}\,\sum_f e_f^2\int_{z_{f}}^{1/2}
dz\,z^3(1-z)^3\,\phi_0^2
\eeeq
where $f$ denotes quark flavours, $m_f$ is quark mass and the diffractive slope $B_d$ in the denominator results from
the $t$-integration of  the structure functions, assuming an exponential form for this dependence.
From HERA data, $B_d=6~\gev^{-2}$.
The variables
\be 
z_{f} = {\textstyle{\frac{1}{2}}}(1-\sqrt{1-4m_f^2/M^2})\,,~~~~
Q^2_f=z(1-z)Q^2+m_f^2
\ee 
and the functions  $\phi_i$  take the following form for $i=0,1$
\be\label{eq:8}
\phi_i
=\int_0^\infty dr r K_i\!\left(Q_f r\right)
J_i\!\left(k_fr\right) \sigmahat(\xp,r)
\ee
where $k_f=\sqrt{z(1-z)M^2-m_f^2}$ is the quark transverse momentum while
$K_i$ and $J_i$ are the Bessel functions.
The lower integration limit $z_f$ in Eqs.~(\ref{eq:5a}) and (\ref{eq:5}) corresponds to a minimal value
of $z$ for which the diffractive  state with mass $M$ can be produced. In such a case $k_f=0$. 
At the threshold for the massive quark production  $M^2=4m_f^2$ and $z_f=1/2$, leading to $F_{T,L}^{(q\qbar)}=0$.
For massless quarks $z_f=0$.

The quantity $\sigmahat(\xp,r)$ in Eq.~(\ref{eq:8}) is called a dipole 
cross section  and described the interaction of the $q\qbar$ dipole with the proton. It
brings the energy dependence into the structure function formulae and is related to the imaginary part of the dipole scattering amplitude, $N(\xp,r,b)$, by the integral over the impact parameter
\be\label{eq:csdipolowy}
\sigmahat(\xp,r)=2\int d^2b\,N(\xp,r,b)\,.
\ee
Notice that for DDIS the Bjorken variable $x$ is substituted by $\xp=x/\beta$.
For $\beta\sim 1$ this  substitution is subleading from the point of view
leading logarithms of energy $W$ which appear in the QCD computation
of this amplitude. However, for large diffractive masses, $\beta\ll 1$, 
such a  substitution becomes phenomenologically important. 

\begin{figure}[b]
\begin{center}
\includegraphics[width=8cm]{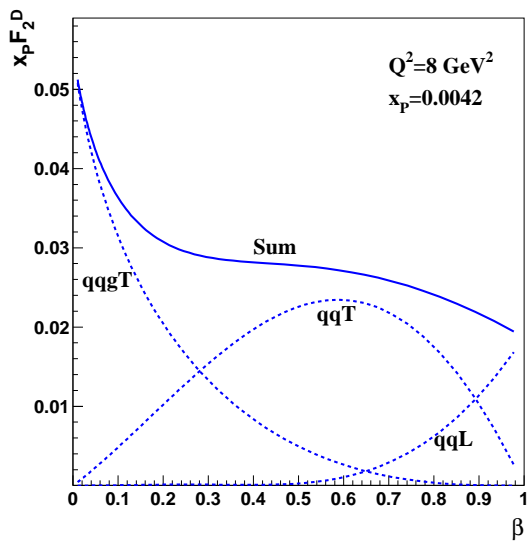}
\caption{The three  components in Eq.~(\ref{eq:1})
as a function of $\beta$ in the massless quark limit 
with the GBW parameterisation of the dipole cross section. 
The $q\qbar gT$ component is without the color factor modification.}
\label{fig:0}
\end{center}
\end{figure}

The $q\qbar g$ diffractive component from transverse photons, computed for massless quarks is given by
\beeq\nonumber
\xp F_T^{(q\qbar g)}
\eq
\frac{81 \beta\alpha_s }{512\pi^5 B_d}\;
\sum_f e_f^2 
\int_\beta^1 \frac{dz}{(1-z)^3} 
\\\nonumber
&\times&\left[\left(1-\frac{\beta}{z}\right)^2+\left(\frac{\beta}{z}\right)^2\right]
\\\nonumber
\\\label{eq:10}
&\times&
\int_0^{(1-z)Q^2} dk^2 \log\left(\frac{(1-z)Q^2}{k^2}\right)\phi_2^2
\eeeq
where the function $\phi_{2}$ takes to form
\be\label{eq:11}
\phi_{2}
=k^2
\int_0^\infty dr\, r\, K_{2}\!\left(\sqrt{\frac{z}{1-z}}kr\right)\,
J_{2}(kr)\,  \hat{\sigma}(\xp,r)
\ee
with $K_{2}$ and $J_{2}$ are the Bessel functions. 
In papers \cite{Wusthoff:1997fz,Golec-Biernat:1999qd}, formula (\ref{eq:10}) was computed with two gluons exchanged between the diffractive system and the proton.
Then, the two gluon exchange interaction was substituted by the
dipole cross section $\hat\sigma=\hat{\sigma}(\xp,r)$ for the $q\qbar$ dipole interaction with the proton. 
For example, for the GBW parameterisation of the dipole cross section \cite{Wusthoff:1997fz}, which we discuss in the next section, is given by
\be
\sigmahat\equiv\sigmahat_{q\qbar}=\sigma_0\left(1-\eto^{-r^2Q_s^2/4}\right)\,.
\ee
However, the $q\qbar g$ system was computed in the approximation when parton transverse momenta fulfil the condition $k_{Tq}\approx k_{T\qbar}\gg k_{Tg}$.
Thus, in the large $N_c$ approximation, it can be treated as a gluonic
color dipole $gg$.  Such a  dipole interacts with  the relative color factor $C_A/C_F$ with respect to  the $q\qbar$ dipole.
Therefore,  the two gluon exchange formula should be  eikonalized with this color factor absorbed into the exponent. For the GBW parameterisation, this  leads to the following gluon dipole cross section in Eq.~(\ref{eq:11})
\be\label{eq:sgg}
\sigmahat\equiv\sigmahat_{gg}=\sigma_0\left(1-\eto^{-(C_A/C_F) r^2Q_s^2/4}\right)\,.
\ee
In such a case, the color factor $C_A/C_F=9/4$ (for $N_c=3$) disappears from the normalisation of the scattering amplitude and we have to rescale the structure function in the following way
\be\label{eq:sgga}
F_T^{(q\qbar g)}~\to~\frac{1}{(C_A/C_F)^2}\,F_T^{(q\qbar g)}\,.
\ee
By the comparison with HERA data, 
we will show in the next section that the latter possibility is more appropriate
for the data description.

We summarise our considerations in Fig.~\ref{fig:0}, which shows three 
components of $F_2^D$ as a function of $\beta$ for fixed values of $\xp$ and $Q^2$.
Each component dominates in different regions of diffractive mass: $F_T^{(q\qbar)}$  
dominates 
for $M^2\sim Q^2$ ($\beta\sim 1/2$), $F_L^{(q\qbar)}$ is important for $M^2\ll Q^2$ ($\beta\approx  1$) and
$F_T^{(q\qbar g)}$ wins for large diffractive mass, $M^2\gg Q^2$ ($\beta\ll 1$).

\section{Dipole scattering amplitude}
\label{sec:dipscatamp}

We are going to compare the presented dipole description of the diffractive structure
functions with the newest HERA data. For this purpose, we consider two  parameterisations of the dipole cross section which are based on the idea of parton saturation in  dense gluon systems. The first one is the GBW parameterisation
with heavy quarks \cite{Golec-Biernat:1998js} which has played an inspirational role in studies of parton saturation in the recent ten years. The  second one is the CGC parameterisation \cite{Iancu:2003ge,Soyez:2007kg} which somehow summarizes the studies within the Color Glass Condensate \cite{Iancu:2003xm} approach to parton saturation. Quite surprisingly, these two parameterisations
give very similar results for the diffractive structure functions. The main reason is the same normalisation of the dipole cross section, $\sigma_0$. The origin of the same numerical value, however, is different. For the GBW parameterisation 
$\sigma_0$ is fitted to the data for $F_2$ while for the CGC parameterisation it is computed from a diffractive slope $B_D$, see Eq.~(\ref{eq:20a}). 

The two considered parameterisations, specified below, describe very well the inclusive
DIS data on the structure function $F_2$. Their use for the DDIS description is a very
important test of the universality of the dipole approach to DIS diffraction.

\begin{itemize} 
\item[(1)] {it  The GBW parameterisation} with heavy quarks has the following form of the $q\qbar$ dipole cross section \cite{Golec-Biernat:1998js}
\be
\sigmahat(\xp,r)=\sigma_0\left(1-\exp(-r^2Q_s^2/4\right))
\ee
where $\sigma_0=29~{\textrm mb}$, and the saturation scale  is given by
\be
Q_s^2=\left({\xp}/{x_0}\right)^{-\lambda}~\gev^2
\ee
with $x_0=4\cdot 10^{-5}$ and  $\lambda=0.288$.
The dipole scattering amplitude in such a  case reads
\be
\hat{N}(\xp,\rbo,\bbo)=\theta(b_0-b)\left(1-\exp(-r^2Q_s^2/4\right)
\ee
where $2\pi b_0^2=\sigma_0$. This form corresponds to a model of the proton with a sharp edge.

\item[(2)] {\it The CGC parameterisation}  with heavy quarks of
the quark dipole scattering amplitude is given by \cite{Iancu:2003ge,Soyez:2007kg,Marquet:2007nf} 
\be\label{eq:19}
\hat{N}(\xp,\rbo,\bbo)=S(\bbo)\,N(\xp,\rbo)\,.
\ee
where the form factor 
$
S(\bbo)=\exp(-b^2/(2B_d))
$
with the diffractive slope from HERA, $B_d=6~\gev^{-2}$.
Thus, the dipole cross section (\ref{eq:csdipolowy}) is given by the formula
\be\label{eq:20}
\sigmahat(\xp,\rbo)=4\pi B_d\, N(\xp,\rbo)\,.
\ee
We see that the asymptotic value of $\sigmahat$ for $r\to \infty$ is the same as for the GBW parameterisation,
if the diffractive slope measured at HERA is substituted, 
\be\label{eq:20a}
\sigma_0=4\pi B_d=29~{\textrm mb}\,.
\ee
In addition,
\beeq\label{eq:21}
&&N(\xp,\rbo)=
\\\nonumber
&&\left\{ 
\begin{array}{ll}
N_0\left(\frac{rQ_s}{2}\right)^{2\gamma_s} \eto^{\frac{2\ln^2(rQ_s/2)}{\kappa\lambda\ln(\xp)}} &     
\mbox{\textrm for~~~ $rQ_s\le 2$} \\
1-  \eto^{-4\alpha\ln^2(\beta rQ_s)} &     \mbox{\textrm for~~~ $rQ_s > 2$}
\end{array}
\right.
\eeeq
where the saturation scale $Q_s$ has now the following parameters: $\lambda=0.22$ and $x_0=1.63\cdot 10^{-5}$.
The parameters  $\alpha=0.615$ and $\beta=1.006$ are chosen such that $N$ and its first
derivative are continues at the point $r$ where $N(r)=N_0=0.7$. The remaining parameters  are given by $\kappa=9.9$ and $\gamma_c=0.7376$.
\end{itemize}

Both parameterisations provide the energy dependence of the diffractive structure
function through the variable $\xp$. This dependence is
determined from fits of the dipole model formula for $F_2$ into the data from HERA 
for the Bjorken variable $x\le 0.01$. In the case of DDIS, $x$ is substituted by $\xp$. 

\begin{figure}[t]
\begin{center}
\includegraphics[width=8cm]{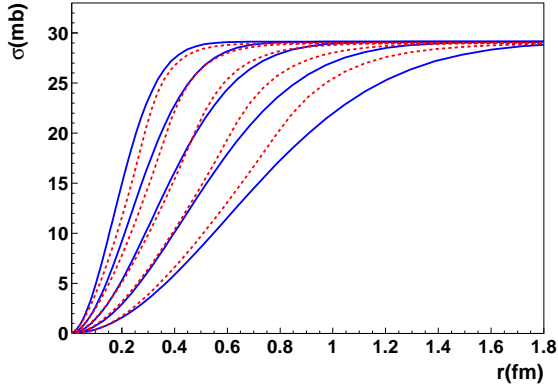}
\caption{The dipole cross section as a function of $r$  for  
$x=10^{-2}\ldots 10^{-6}$ (from right to left) and for the GBW (continuous lines) and CGC (dashed lines)  parameterisations.}
\label{fig:0a}
\end{center}
\end{figure}

\section{Diffractive charm quark production}
\label{sec:charm}

\begin{figure}[t]
\begin{center}
\includegraphics[width=6.5cm]{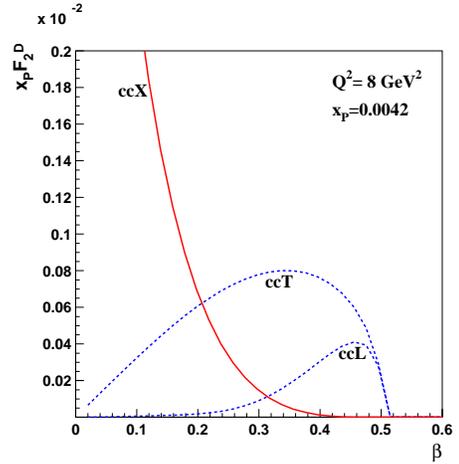}
\includegraphics[width=6.5cm]{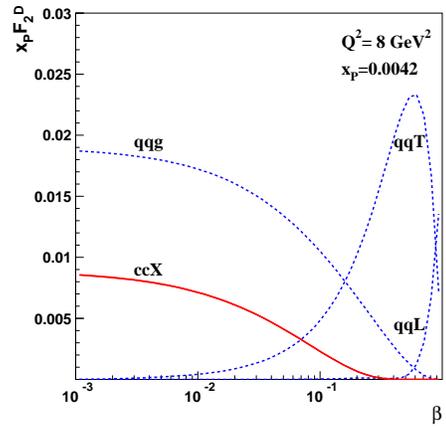}
\caption{
Top: the $c\cbar T$ and $c\cbar L$ components of $F_2^D$ from the dipole model with the GBW parameterisation
together with the $c\cbar X$ contribution from the collinear factorisation approach (\ref{eq:13})
 with the  diffractive gluon distribution (\ref{eq:17d}). Bottom: the $c\cbar X$ component 
in a different scale against the massless $q\qbar T$, $q\qbar L$ and $q\qbar g$ components.
}
\label{fig:4}
\end{center}
\end{figure}

\begin{figure*}[t]
\begin{center}
\epsfig{figure=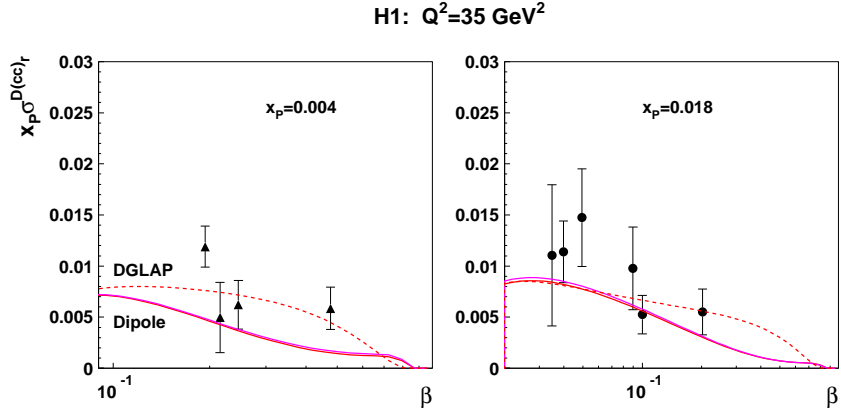,width=12cm}
\caption{A comparison of the collinear factorisation predictions with the GBW and
CGC gluon distributions (solid lines) with the HERA data on the open diffractive charm production.
The dashed lines are computed with the  gluon distribution obtained in the  DGLAP fit \cite{GolecBiernat:2007kv} to the H1 data on the diffractive structure functions.}
\label{fig:6}
\end{center}
\end{figure*}

In the diffractive scattering heavy quarks are produced 
in quark-antiquark pairs, $c\cbar$ and $b\bbar$ for charm and bottom, respectively. 
Such pairs can be produced provided that 
the diffractive mass of is above the quark pair production threshold
\be\label{eq:12}
M^2=Q^2\left(\frac{1}{\beta}-1\right) > 4m_{c,b}^2
\ee
In the lowest order the diffractive state consist only the  $c\cbar$ or $b\bbar$ pair.
In the forthcoming we consider only charm production since bottom production is negligible.
The corresponding contributions to $F_2^D$ are given by Eqs.~(\ref{eq:5a}) and 
(\ref{eq:5}) with one flavour component. For example, for charm production from transverse photons we have 
\beeq\nonumber
\label{eq:5new}
\xp F_T^{(c\cbar)}\!\!\eq\!\!
\frac{3 Q^4 e_c^2}{64\/\pi^4\beta B_d} \int_{z_{c}}^{1/2}
dz\, z(1-z)
\\
&\times&\left\{
[z^2+(1-z)^2]\,Q^2_c\,\phi_1^2 \,+\,m_c^2\, \phi_0^2
\right\}
\eeeq
where $m_c$ and $e_c$ are charm quark mass and electric charge, respectively.
The minimal value of diffractive mass equals
$M^2_{min}=4m_c^2$, thus the maximal value of $\beta$ is given by
\be
\beta_{max}=\frac{Q^2}{Q^2+4m_c^2}\,.
\ee
In such a case,  $z_c=1/2$ in Eq.~(\ref{eq:5new})
and  $F_{T,L}^{(c\cbar)}=0$ for $\beta> \beta_{max}$. This is shown in Fig.~\ref{fig:4} (top)
for the $c\cbar$ diffractive states from transverse $(c\cbar T)$ and longitudinal $(c\cbar L)$
 photons. By the comparison with the corresponding curves 
for three massless quarks $(q\qbar T,q\qbar L)$, shown in Fig.~\ref{fig:4} (bottom), we see that the 
exclusive diffractive charm production  contributes only $1/30$ to the total structure function $F_2^D$. Thus it can practically be neglected.

The next component is the $c\cbar g$ 
diffractive state. Unfortunately, formula (\ref{eq:10}) for the $q\qbar g$ 
production is only known in the massless quark case and  cannot be used for heavy quarks.
Thus, we have to resort to the collinear factorisation formula, given by Eq.~(\ref{eq:13}),
in which the charm-anticharm  pair
is produced via the photon-gluon fusion: $\gamma^*g\to c\cbar$ \cite{Goncalves:2004dv}. If such an approach is
applied to diffractive scattering,  gluon is a ``constituent of a pomeron''. The diffractive state consists 
of additional particles $X$ (called ``pomeron remnant'') 
in addition to the heavy quark pair,  which is well separated in rapidity from the scattered  proton.
The collinear factorisation formula for the charm contribution to the diffractive structure functions is 
taken from the fully inclusive case \cite{Gluck:1994uf} in which the standard gluon distribution is replaced by the diffractive gluon distribution $g^D$:
\beeq
\nonumber
\label{eq:13}
\xp F_{2,L}^{D(c\overline{c}X)}\!\!\eq\!\!
2\beta\, e_c^2\, \frac{\alpha_s(\mu_c^2)}{2 \pi}
\int_{a\beta}^1\frac{dz}{z}\,
C_{2,L}\!\left( \frac{\beta}{z},{\frac{m_c^2}{Q^2}} \right) 
\\
&\times&\xp g^D(\xp,z,\mu_c^2)
\eeeq
where $a = 1 + 4 m_c^2/Q^2$ and the factorisation scale $\mu_c^2=4 m_c^2$
with the charm quark mass  $m_c=1.4~\gev$.  The leading order coefficient functions are given by
\beeq\nonumber
C_2(z, r)\!\! \eq\!\!  \half\left\{z^2 + (1-z)^2 + 4z(1-3z) r -
8z^2 r^2\right\} 
\\\nonumber
&\times& \ln{\frac{1+\alpha}{1 -\alpha}} 
\,+\,\half \alpha\big\{ -1 + 8z(1-z)
\\
&-& 4z(1-z)r \big\}
\\ \nonumber
\\\label{eq:14b}
C_L(z, r)\!\!  \eq\!\!  - 4z^2 r \ln{{1+\alpha}{1 -\alpha}} + 2\alpha z (1 - z)
\eeeq
where $r={m_c^2}/{Q^2}$ and $\alpha=\sqrt{1 -{4 r z}/{(1-z)}}$.
The lower integration limit in Eq.~(\ref{eq:13}) results from the condition
for the heavy quark production in the fusion:  $\gamma^*g\to c\cbar$,
\be
(z\xp\, p +q)^2\ge 4m_c^2
\ee
where we assume that gluon carries a fraction $z$ of the pomeron momentum $\xp p$.

\begin{figure*}[t]
\begin{center}
\includegraphics[width=12cm]{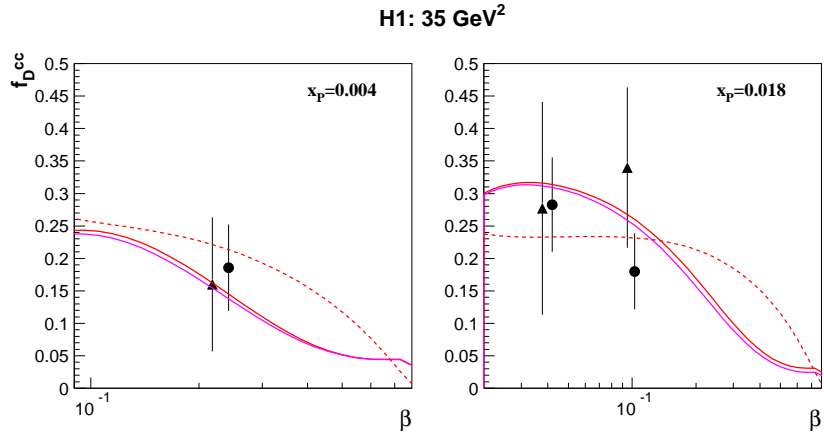}
\caption{
The fractional charm contribution, $f_D^{c\cbar}$ given by Eq.~(\ref{eq:fraccc}), is shown as a function of $\beta$, for two values of $\xp=0.004$ and $0.018$. The solid lines
are computed for  the $c\cbar X$ contribution  with the GBW and CGC diffractive gluon distributions while the dashed lines are found for the  diffractive gluon distribution obtained in the  DGLAP fit \cite{GolecBiernat:2007kv} to the H1 Collaboration data.}
\label{fig:7}
\end{center}
\end{figure*}

The $c\cbar X$ contribution given by Eq.~(\ref{eq:13}) is shown in  Fig.~\ref{fig:4} as the solid lines. 
As seen in the top figure, this component becomes significant for $\beta<0.1$ . By a comparison with the massless quark contributions (the bottom figure) we see that  diffractive charm production contributes up to $30\%$ to the diffractive structure function $F_2^D$ for small values of $\beta$. The presented results were obtained assuming the diffractive
gluon distribution which results from the dipole models, given by Eq.~(\ref{eq:17d}) in Appendix, with the GBW
parameterisation of the dipole cross section with the color factor modification (\ref{eq:colfac}). The CGC parameterisation gives a similar result.

In Fig.~\ref{fig:6} we show the collinear factorisation predictions for the diffractive charm production confronted with the new HERA data \cite{Aktas:2006up} on the charm component of the reduced cross section:
\be
\sigma_r^{D(c\cbar)}=F_2^{D(c\cbar)}-\frac{y^2}{1+(1-y)^2}F_L^{D(c\cbar)}\,.
\ee
The solid curves, which are barley distinguishable, correspond to the 
result with the GBW and CGC parameterisations of the diffractive gluon distributions.
The dashed lines are computed for the gluon distribution from a fit  to the H1 data \cite{GolecBiernat:2007kv} based on the DGLAP equations. 
The present accuracy of the charm data does not allow to discriminate between these two approaches although
the data seem to prefer the gluon distribution from the DGLAP fit which is much more
concentrated in the large $z$-region as compared to the dipole model gluon distributions,
see Fig.~\ref{fig:2qg} in Appendix.

The importance of diffractive charm is illustrated in Fig.~\ref{fig:7} where the fractional charm contribution,
\be\label{eq:fraccc}
f_D^{c\cbar}={\sigma_r^{D(c\cbar)}}/{\sigma_r^{D}}\,,
\ee
to the total diffractive cross section, discussed in the next section,
is shown as a function of $\beta$ against the H1 Collaboration data \cite{Aktas:2006up}.
For small values of $\beta$, the charm contribution equals on average approximately
$20-30\%$, which is comparable to the charm fraction in the inclusive cross section  for similar values of $Q^2$ \cite{Aktas:2005iw}.

\section{Comparison with the HERA data}
\label{sec:f2d}

\begin{figure*}[t]
\begin{center}
\includegraphics[width=13cm]{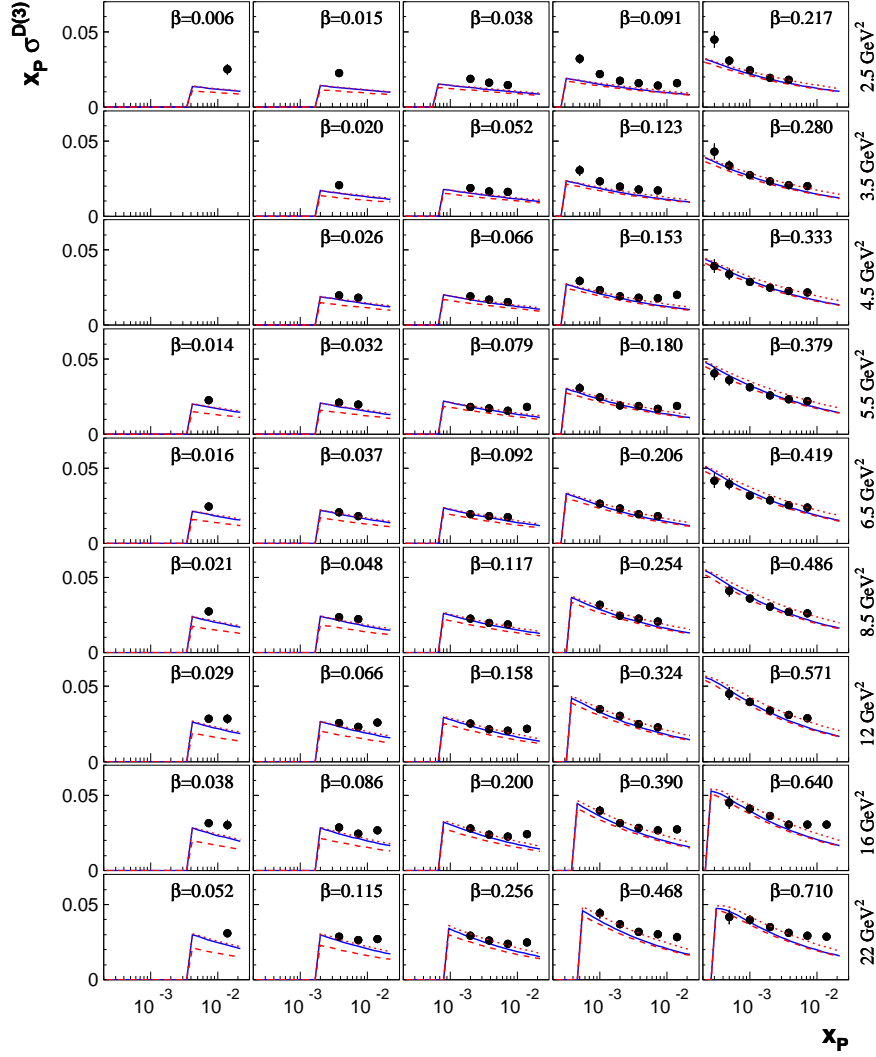}
\caption{A comparison of $\sigma_r^D$ from the two considered dipole models with the newest ZEUS Collaboration data \cite{Chekanov:2008fh}. The solid lines correspond to the GBW parameterisation of the dipole cross section  with the color factor modifications (\ref{eq:sgg}) and (\ref{eq:sgga}), while the dotted lines correspond to the CGC parameterisation.  The dashed lines show the results without the charm contribution.}
\label{fig:4ZEUSa}
\end{center}
\end{figure*}

\begin{figure*}[t]
\begin{center}
\includegraphics[width=13cm]{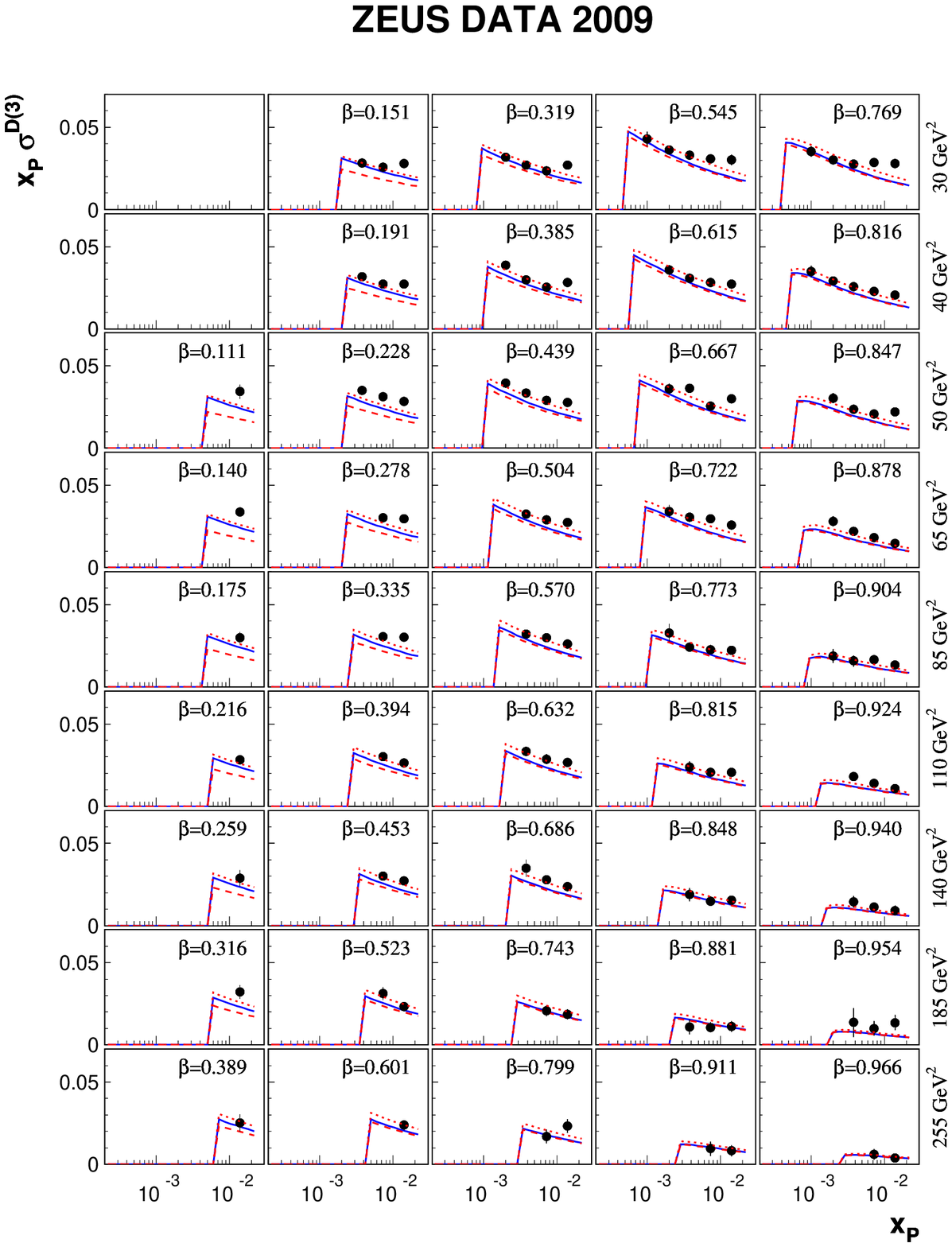}
\caption{The same as in Fig.~\ref{fig:4ZEUSa} but for higher values of $Q^2$. The dashed lines show the contribution without charm. }
\label{fig:4ZEUSb}
\end{center}
\end{figure*}

In Figs.~\ref{fig:4ZEUSa} and \ref{fig:4ZEUSb}
we show a comparison of the dipole
model predictions with the ZEUS Collaboration data \cite{Chekanov:2008fh} on  the reduced cross section
\be
\sigma_r^{D}=F_2^{D}-\frac{y^2}{1+(1-y)^2}F_L^{D}\,.
\ee
We included  the charm contribution in the above structure functions.
The solid lines correspond to the GBW parameterisation of the dipole cross section with the color factor modifications (\ref{eq:sgg}) and (\ref{eq:sgga}) of the $q\qbar g$ component, while the dashed lines are obtained from the CGC parameterisation. We see that the two sets of curves are barely distinguishable. This somewhat surprising results could
be attributed to the same normalisation of the dipole cross section in both models, $\sigma_0=29~{\rm mb}$. Let us emphasise again that this numerical value was obtained in two  different ways (see Sec.~\ref{sec:dipscatamp} for more details).
The color factor modification of the $q\qbar g$ component in the GBW parameterisation is necessary since the curves without such a modification significantly overshoot the data
(by a factor of two or so)
in the region of small $\beta$ where the ${q\qbar g}$ component dominates. 

\begin{figure*}[t]
\begin{center}
\includegraphics[width=13cm]{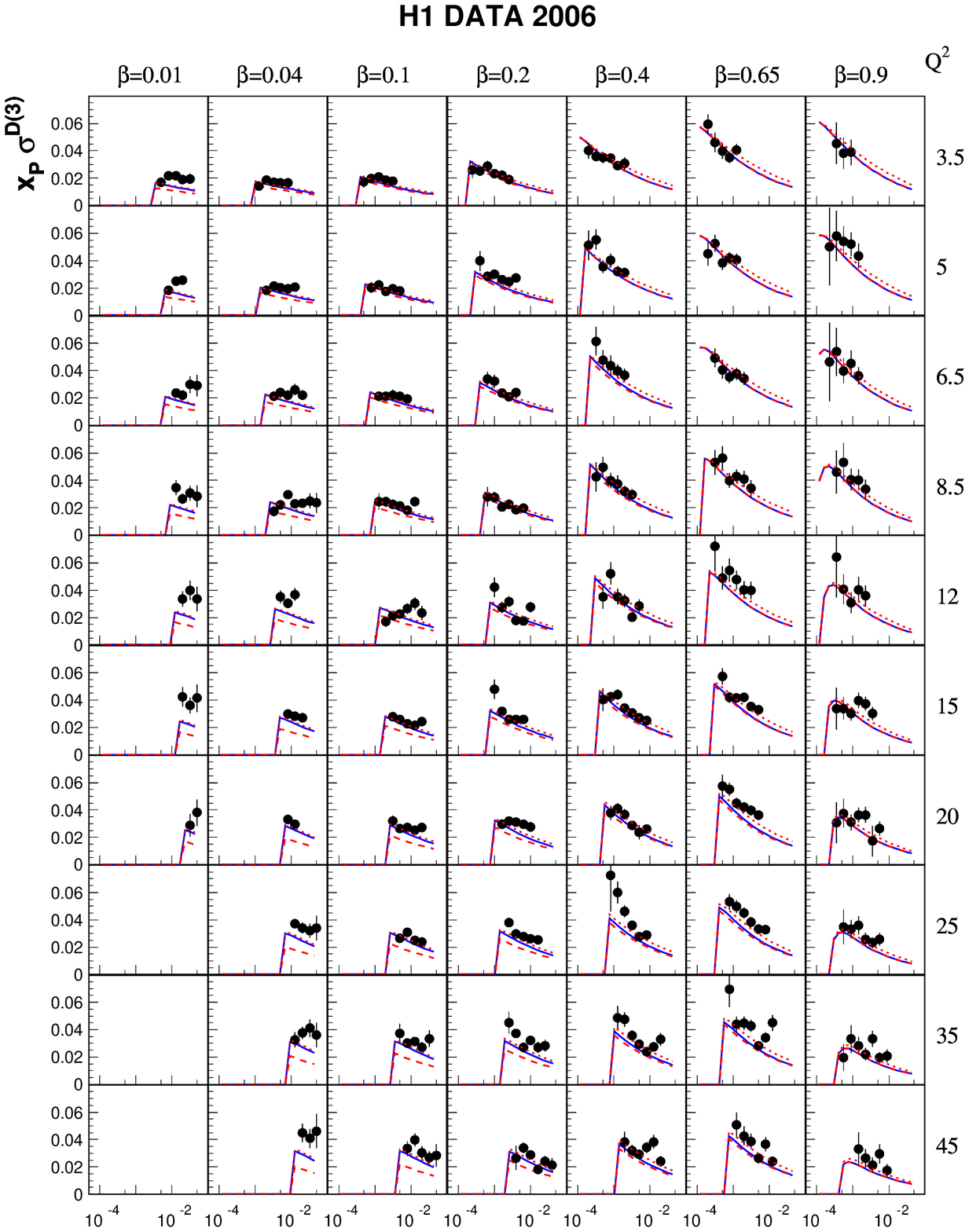}
\caption{The same as in Fig.~\ref{fig:4ZEUSa} but for the H1 Collaboration data \cite{Aktas:2006hy} . The dashed lines show the contribution without charm.}
\label{fig:5H1}
\end{center}
\end{figure*}

The comparison  of the predictions with the data also reveals a very important aspect of  the three component dipole model  (\ref{eq:1}). In the small $\beta$ region, the  curves are systematically  below the data points, which effect  may be attributed to the lack of higher order components in the diffractive state, 
i.e. with more than one gluon or $q\qbar$ pair.  This is also seen for the H1 Collaboration data \cite{Aktas:2006hy} shown in Fig.~\ref{fig:5H1}. For small values of $\beta$ both the solid (GBW) and dashed (CGC) curves
are below the data. It is also important that the charm contribution, described in Sec.~\ref{sec:charm}, is added into the analysis. Without this contribution the comparison would be much worse than that shown here.

\begin{figure*}[t]
\begin{center}
\includegraphics[width=13cm]{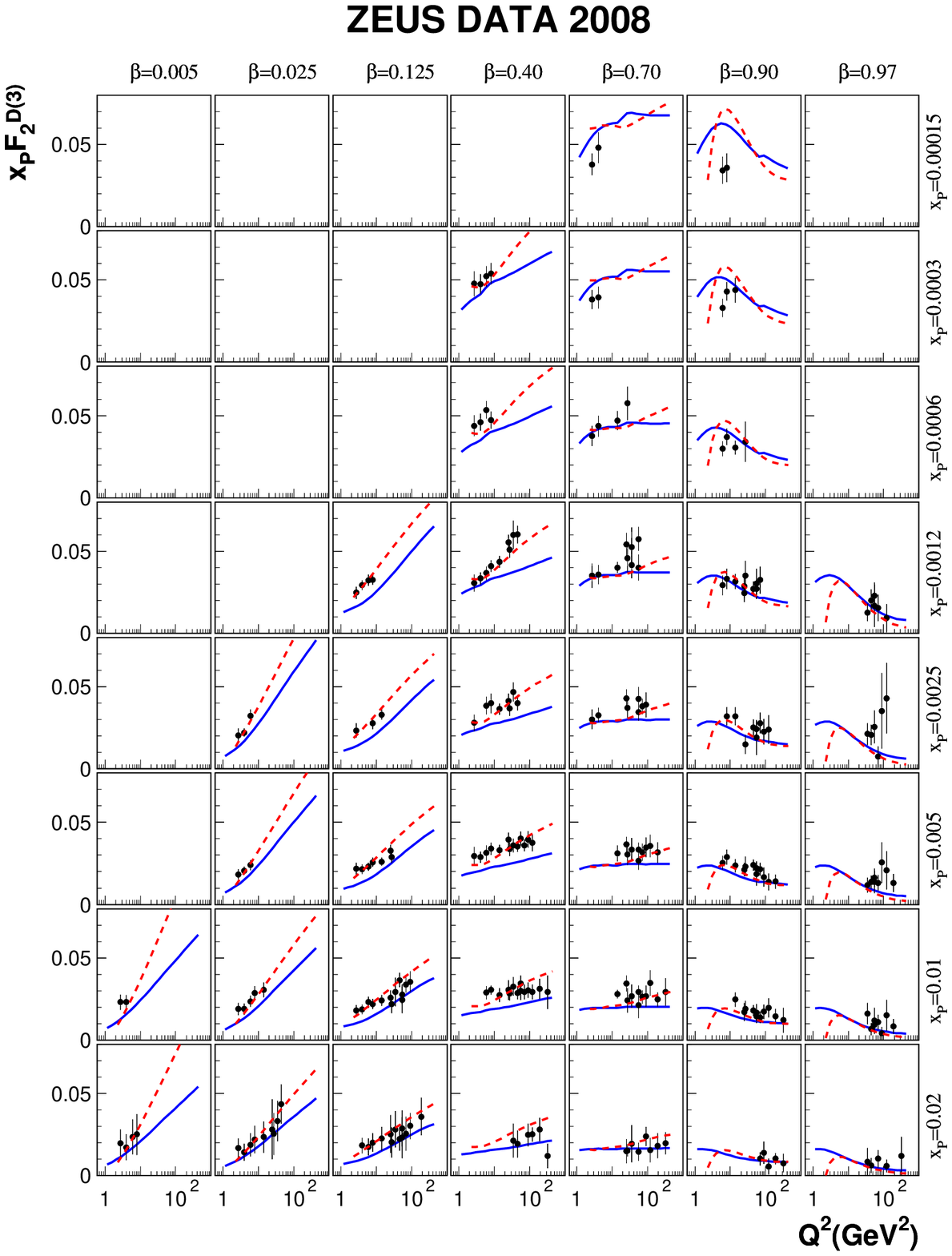}
\caption{A comparison of the GBW dipole model predictions (solid lines) and the results from the DGLAP fit (dashed lines) \cite{GolecBiernat:2007kv} to the ZEUS data \cite{Chekanov:2008cw}.}
\label{fig:7Ze}
\end{center}
\end{figure*}

This effect 
may be attributed to the lack of higher order components in the diffractive state, 
i.e. with more than one gluon or $q\qbar$ pair.  
They may be added in the DGLAP based approach to inclusive diffraction which sums additional  partonic emissions in the diffractive state in the transverse momentum ordering approximation.
A comprehensive discussion of the DGLAP based fits to the diffractive HERA data
is presented in \cite{GolecBiernat:2007kv}. We only recall here that
in this approach the diffractive structure functions are twist-$2$ quantities with
the logarithmic dependence on $Q^2$ for fixed $\xp$ and $\beta$. They are related
to the diffractive parton distributions by the standard collinear factorisation formulae,
e.g. in the leading $\log Q^2$ approximation we have:
\be\label{eq:drp1}
F_2^D(\xp,\beta,Q^2)=\sum_f e_f^2\,\beta\,(q^D_f+\overline{q}^D_f)\,,
\ee
where $q_f^D$ and $\overline{q}^D_f$ are diffractive quark/antiquark distributions.
We additionally assume  flavour democracy for these distributions to account for
vacuum quantum number exchange responsible for diffraction, 
\be\label{eq:drp1a}
q^D_f=\overline{q}^D_f=\frac{1}{2N_f}\Sigma^D
\ee
where $\Sigma^D$ is a diffractive singlet quark distribution. 
This distribution is  evolved in $Q^2$ by the DGLAP equations
together with the gluon distribution $g^D$.
In contrast to the dipole model case,
the $\xp$ dependence of the diffractive parton distributions is fitted to data,  as well as their form in $\beta$ at some initial scale $Q_0^2$. In Fig.~\ref{fig:7Ze} we show the result of such an analysis  (dashed lines) applied to the ZEUS data \cite{Chekanov:2008cw}. In the  
small$-\beta$ region, the DGLAP fit curves are going through the experimental points with larger
logarithmic slope, $\partial F_2^D/\partial \ln Q^2$, than in the dipole approach. 
This illustrates the importance of  more complicated diffractive states than the $q\qbar g$ state.

\section{Conclusions}

We presented a comparison of the dipole model results on the diffractive structure functions with the HERA data. We considered two most popular parameterisations of the interaction between the diffractive system and the proton
(the GBW and CGC parameterisations) which are based on the idea of parton saturation.
The three component model with the $q\qbar$ and $q\qbar g$ diffractive states  describe reasonable well  the recent data. 
However,  the region of small values of $\beta$ needs some refinement by considering components with more gluons and $q\qbar$ pairs in the diffractive state. This can be achieved in the DGLAP based approach which sums partonic emissions in the diffractive state in the transverse momentum ordering approximation. We extracted the diffractive gluon distribution from the  dipole model formulae
to use it for the computation of the charm contribution to $F_2^D$. 
We found good agreement with the HERA data on the diffractive open  charm production both for the
the gluon distributions from the considered dipole models and the DGLAP fits from \cite{GolecBiernat:2007kv}.
The latter statement, however, is possible to make only due to present accuracy of the  charm data. 
The results presented in this work might be a starting point for the future electron-proton collider LHeC
at CERN.

\begin{acknowledgments}
This work is partially supported by the grants of MNiSW Nos. N202 246635 and
N202 249235 and the grant HEPTOOLS, MRTN-CT-2006-035505.
\end{acknowledgments}

\appendix*
\section{Diffractive gluon distribution from dipole models}

\begin{figure}[t]
\begin{center}
\includegraphics[width=7cm]{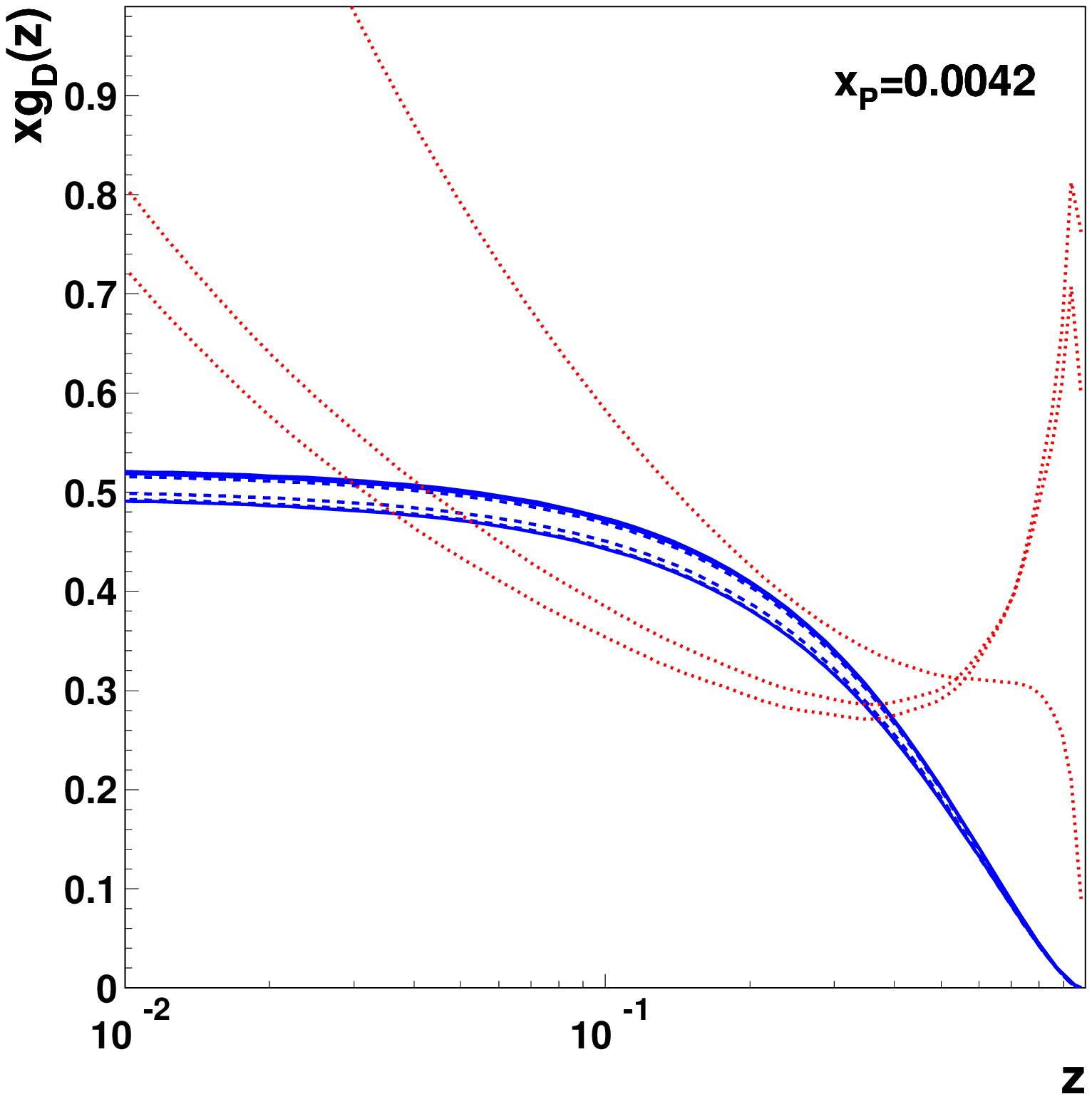}
\caption{
The diffractive gluon distribution $xg_D(\xp,z,Q^2)$
for $Q^2=4m_c^2,\,7.84,\, 10,\, 100~\mbox{GeV}^2$ (from bottom to top) 
and $\xp=0.0042$. The solid lines correspond to the GBW parameterisation while
the dashed lines to the  CGC parameterisation. The dotted lines show the 
gluon distributions from the DGLAP fit \cite{GolecBiernat:2007kv} to the H1 Collaboration data.
}
\label{fig:2qg}
\end{center}
\end{figure}

A comprehensive discussion of the derivation of the diffractive parton distributions in dipole models
can be found in \cite{Golec-Biernat:2001mm}. Here we only recall the derivation of 
the diffractive gluon distribution $g^D(\xp,z,Q^2)$ which supplements that in \cite{Golec-Biernat:2001mm}.
We start from Eq.~(\ref{eq:10}) which we  reduce to the collinear factorisation form. 
Let us substitute
$
(1-z)Q^2\to Q^2
$
in there. We numerically checked that such a substitution practically does not change the diffractive structure 
function. Then  the logarithmic derivative of $F_T^{(q\qbar g)}$ reads
\beeq\nonumber
\frac{\partial F_T^{(q\qbar g)}}{\partial \ln Q^2}
\!\!\eq\!\!
\frac{81 \beta\alpha_s }{512\pi^5\xp B_d}\,\sum_f e_f^2 
\int_\beta^1 \frac{dz}{z} 
\left[\left(1-\frac{\beta}{z}\right)^2+\right.
\\\label{eq:ap2}
&+&\left.\left(\frac{\beta}{z}\right)^2\right]
\frac{z}{(1-z)^3}\int_0^{Q^2} dk^2
\phi_2^2\,.
\eeeq
On the other hand, from the DGLAP
evolution equation  we have for the diffractive singlet quark distribution (\ref{eq:drp1a})
\beeq\nonumber
\frac{\partial \Sigma^D}{\partial \ln Q^2}\!\!&\approx&\!\! \frac{\alpha_s(Q^2)}{2\pi}
\int_\beta^1\frac{dz}{z}\,N_f\left\{\left(1-\frac{\beta}{z}\right)^2
+\left(\frac{\beta}{z}\right)^2\right\}
\\
&\times& g^D(\xp,z,Q^2)
\label{eq:13d}
\eeeq
where we neglected on the right hand side a contribution with the singlet quark distribution which is much smaller
than the gluonic one when $\beta\ll 1$. Thus from Eq.~(\ref{eq:drp1}) we find for the diffractive structure function 
\beeq\nonumber
\label{eq:14d}
\frac{\partial F_2^D}{\partial \ln Q^2}\!\!\eq\!\!
\frac{\beta\alpha_s}{2\pi}\,\sum_f e_f^2
\int_\beta^1\frac{dz}{z}\left\{\left(1-\frac{\beta}{z}\right)^2
+\left(\frac{\beta}{z}\right)^2\right\}
\\
&\times&g^D(\xp,z,Q^2)\,.
\eeeq
For small  $\beta$ we have: $F_T^{(q\qbar g)}\approx F_2^D$, thus by the comparison with 
Eq.~(\ref{eq:ap2}), we find the following diffractive gluon distribution
\be
\label{eq:17d}
g^D(\xp,z,Q^2)=\frac{81}{256\pi^4 \xp B_d }\frac{z}{(1-z)^3}\int_{0}^{Q^2}dk^2\,
\phi_2^2
\ee 
where
\be\label{eq:17e}
\phi_{2}=k^2
\int_0^\infty dr\, r\, K_{2}\!\left(\sqrt{{z}/{(1-z)}}kr\right)\,
J_{2}(kr)\,  \hat{\sigma}(\xp,r)\,.
\ee
For the GBW parameterisation of the dipole cross section,
we additionally rescale the gluon distribution,
\be\label{eq:colfac}
g^D~\to~ \frac{1}{(C_A/C_F)^2}\,\,g^D\,,
\ee
and use formula (\ref{eq:sgg}) for the dipole cross section.
For the CGC parameterisation this rescaling has been already taken into account.
In Fig.~\ref{fig:2qg}  we show the gluon distributions computed for the GBW parameterisation with the color factor modification (solid lines) and for the CGC  parameterisation (dashed lines).
There is practically no difference between them for the indicated scales.  
For $Q^2>4m_c^2$, the $Q^2$ dependence  of the gluon distribution (\ref{eq:17d})  is already very weak and  close to the asymptotic limit obtained for $Q^2\to \infty$. 
We also show in this figure the gluon distributions
found in a DGLAP fit with higher twist to the recent H1 data  \cite{GolecBiernat:2007kv} (dotted lines) with a strong dependence on $Q^2$ due to the DGLAP evolution.

\bibliographystyle{h-physrev4}
\bibliography{mybib}

\end{document}